# eHEALTH TECHNOLOGY INTEGRATION WITH HEALTHCARE WORK ACTIVITIES IN PUBLIC HOSPITALS: *A CRITICAL REALIST PERSPECTIVE*

Mourine Achieng, Cape Peninsula University of Technology, sachiengm@gmail.com

Ephias Ruhode, Cape Peninsula University of technology, Ruhodee@cput.ac.za

**Abstract:** Integration of eHealth technologies with healthcare work activities has seen great advancement in many healthcare systems in developing countries. However, these efforts have been tainted by several challenges such as fragmentation, lack of standardization and co-ordination. Subsequently, the undertakings of eHealth articulated in health strategy/policy documents have not been fully realised. The implications of this has been that the majority of the population still access inadequate healthcare services. The aim of this paper is to explain why the current integration efforts do not adequately facilitate healthcare work activities in public hospitals in under-served contexts of South Africa. A critical realist perspective within a qualitative approach was adopted. A total of 21 participants were purposively sampled and interviewed because of their knowledge and experience in the healthcare service delivery process as well as their involvement in the integration of ehealth. The study applied the Activity Analysis and Development (ActAD) model as a theoretical analytical tool and draws on normalization process theory (NPT) as an explanatory framework. The findings highlight generative mechanisms such as the inadequate analysis of system's fit-for-purpose in healthcare workflows have inhibiting effects in the integration process of eHealth.

## 1. INTRODUCTION

Access to inadequate public healthcare service delivery is a constant reality for the majority of the population in developing countries. Inequity in accessing quality healthcare services in developing countries is often attributed to scarce infrastructure resources, shortage in healthcare practitioners, poor leadership and management and so forth (Omotoso & Koch, 2018; Maphumulo & Bhengu, 2019). In addition, challenges such as inadequate data management and communication at different levels of healthcare systems result to delays in the delivery of healthcare services (Nobana et al., 2020). Subsequent to these challenges, governments have turned to eHealth technologies to address the inadequacies in their healthcare systems. The term eHealth can be referred to as the use of information and communication technologies (ICT) as well as information systems (IS) to facilitate healthcare work activities for an improved service delivery (Andreassen et al., 2015). eHealth technologies are often regarded as enablers that offer quality enhancing efficiencies to operational and decision making processes in the healthcare sector. eHealth technologies are also expected to increase healthcare practitioner's productivity and in turn, improve patient experience in the delivery process - by gaining access to timely, cost effective and quality care services (Liu et al., 2013; Negash et al., 2018).

Although there are several benefits associated with ehealth technologies, the implementation and integration of these technologies has always been problematic (Cresswell & Sheikh, 2013; Calligaro et al., 2017). Integration in this paper is defined as the process of putting to use technology interventions within a healthcare setting for the purpose of improving healthcare service delivery.

Cresswell and Sheikh (2013) further postulate that technology integration in healthcare organizations is complex. This can be attributed to the complex nature of healthcare service delivery, coupled with a wide range of inter-related socio-technical factors that are shaped by organisational factors. Failure to adequately identify and address the socio-technical factors further exacerbates the complexities (Farzandipur et al., 2016). The current integration efforts especially in developing countries, have been tainted by several challenges such as fragmentation, lack of standardization, co-ordination and interoperability (Adenuga et al., 2015; Fletcher, 2016; Senyoni, 2020). As such, the integration of eHealth technologies with healthcare work activities has not yielded desired outcomes. The integration of eHealth technologies requires major structural reforms such as implementing policies and strategies that would inform equal funding and infrastructure resource distribution, allow for increased human resource capacity development and many other requirements that would facilitate ehealth integration process. These reforms would result in changes in healthcare setting, practitioner's work practices, patient's service expectations and experiences in the delivery of healthcare service and overall improved health outcomes.

In light of these, the research question was, *why does the current integration of eHealth technologies with healthcare work activities not adequately facilitating public healthcare service delivery?* In addressing this question, the study first determined the context-based factors that influence eHealth integration with healthcare work activities in public hospitals in resource-constrained environments. The study then establishes generative mechanisms in the integration process of eHealth technologies with healthcare work activities that generate current eHealth integration outcomes. Generative mechanisms are described as the "underlying entities, processes or structures which operate in a particular contexts to generate outcomes of interest" (Astbury & Leeuw, 2010, p. 368).

The rest of the paper is structured as follows: section 2 presents a review of eHealth technologies in South Africa. Section 3 presents a discussion on the approach applied in the study, section 4 presents the results of the interview process. Section 5 presents the analysis and discussion on the factors that influence eHealth integration with healthcare work activities and the generative mechanisms in the integration process. In the last section, the study's conclusions are drawn.

## 2. ELECTRONIC HEALTH LANDSCAPE IN SOUTH AFRICA

The majority of the population access healthcare services from a dysfunctional public healthcare system characterised by great disparities in funding provision, under staffing and maldistribution of resources (Pillay, 2001; Versteeg et al., 2013). Consequently, many South Africans who rely on public healthcare services have access to poor quality care services (Govender et al., 2018; Malakoane et al., 2020). Subsequent to this, the government has sought innovative ways of strengthening the public healthcare system through the adoption, implementation and integration of eHealth technologies. The eHealth technologies are classified into three categories: (i) patient-care level systems such as clinical care and supporting services; (ii) operational-level systems that are used for monitoring and evaluation, and for administrative purposes; and (iii) strategic-level systems.

The district health information system (DHIS) has been in use in South Africa at the primary health care level since the mid-1990s and has seen a number of success. However, an increase in demand for routine information in the public healthcare sector exposed gaps in the DHIS system such as poor data quality and dataflow bottlenecks. Another example of the eHealth initiatives is the electronic health records (EHR) systems, they are also commonly used in most public healthcare facilities. The TrakCare Lab system, is commonly used by most laboratories and is responsible for all diagnostic pathology tests in the public healthcare sector. The national Health Patient Registration System (HPRS) implemented in healthcare facilities countrywide are mainly used for monitoring and evaluation. The system allows for patients identity verification and records the purpose of visit, this is a reliable source of national patient demographic data (CSIR, 2016). In recent years, the digital and mobile health platforms in South Africa has seen great initiatives. The use of these platforms has resulted in significant reduction in cost to both patients and healthcare facilities as the Department of health is able to collect real-time data easily (Barron et al., 2018). An example is the MomConnect mobile health initiative that is used to provide pregnant and postpartum women with weekly health information via instant messaging apps (Barron et al., 2018). This initiative is integrated into maternal and child healthcare services.

Despite great achievements that have been realised through the use of eHealth technologies, the public healthcare sector still experiences challenges (Omotoso & Koch, 2018; Maphumulo & Bhengu, 2019). For example, there are still challenges with poor data quality and dataflow bottlenecks (Mchunu, 2013), and reporting discrepancies at different levels.

## 3.    RESEARCH APPROACH

The objective of this paper was primarily to provide explanations for the current outcomes of eHealth integration efforts in public hospitals in under-served contexts. Based on this objective, the critical realist perspective was chosen within which qualitative research methods were employed. The critical realist stance, allows researchers to view the empirical knowledge as one that is socially constructed (Bhaskar, 1979). The critical realist stratified ontology provided the depth of understanding of and explanation for events/effects in the integration of eHealth activities through generative mechanisms. The three-level ontology includes the domain of reality, actual and real (Bhaskar (1986). Semi-structured interviews were conducted with purposively sampled participants because of their experience in the healthcare service delivery process and involvement in the integration process of eHealth technologies. In total there were 21 participants in the study. Of these, five were managers at the hospital in various units. Two of the participants were managers who oversaw eHealth initiatives at provincial level. The remaining number of participant were healthcare practitioners at the hospital who carried out either clinical care activities or administrative activities using eHealth technologies. The objective of the interviews was to determine participant's awareness and use of eHealth technologies and also determine their perceptions of eHealth integration with healthcare work activities and its effect on service delivery.

A theoretical thematic analysis technique was employed for the analyses of the empirical data. This technique allowed the researchers to repeatedly search through the empirical data set in order to identify patterns that emerged. The researchers manually identified, analyzed and recorded the patterns (themes) within the empirical data as illustrated in Tables 1 and 2 in the analysis section.

### 3.1. Case Description

The empirical case, Nelson Mandela Academic Hospital is located in the O.R. Tambo District Municipality in the Eastern Cape province of South Africa. It is one of the largest provincial government funded hospital in the region and serves a number of patients from in and around the municipality. The hospital was primarily selected because it is in a rural or under-served context and also because it makes use of eHealth technologies such as DHIS and HPRS.

### 3.2. Validity and reliability

Validity and reliability are important factors to consider in a research design or in the analysis of the results. The concept of validity is described across qualitative studies by a wide range of terms. Winter (2000, p.1) states that validity is "…a contingent construct, inescapably grounded in the processes and intentions of a particular research methodologies and projects." On the other hand, reliability is viewed as the exact replicability of the results. In qualitative research such as this one, the essence of reliability lies with consistency. In this paper, purposive sampling technique was employed to select participants of the study and as such the outcomes of the study are not generalizable. The constructs of the ActAD model as well as the key variables from the objectives of the study were employed to determine and validate respectively. Since the study is largely subjective, reliability was ensured by operationalizing key variables of the objectives. This technique may not yield the maximum reliability due to the subjective nature of the study.

### 3.3. Theoretical/Analytical Framework

While the paper gives recognition to other social theories and models in the IS field, the ActAD model and NPT theory were adopted as analytical theoretical frameworks. Based on Activity Theory (AT), the ActAD model was employed as an analytical lens. Its constructs include subjects, object, tools, objectives, rules, transformation and outcomes (Korpela et al., 2000). The model also adopts the concept of the work activity system adapted from Engeström's (1987) systematic structure of work activity. The concept views social activities such as eHealth integration as rules-based, deliberate and collective work by various subjects, in pursuit of a common purpose. The object construct is perceived as the purpose of which a social activity is being carried out. Subject construct is viewed as stakeholders in the work activity system who make use of tools (eHealth), policies and procedures (rules) to carry out their work activities. The transformation construct refers to the actual work process, where policies, tools, procedures and activities converge to produce an outcome.

NPT is a theory on the collective 'work or effort' done individually or collectively to implement and sustain an intervention (May, 2006). The enacting of an intervention are done through its four constructs or mechanisms; coherence, cognitive participation, collective action and reflexive monitoring (May & Finch, 2009). Coherence relates to how work activities define and organize interventions in an organizational context are understood, perceived meaningful and the invested in. Cognitive participation relates to the commitment required from individual actors as well as the degree to which those individuals are engaged with the new intervention. Collective action relates to the work that will be required of participants to in the integration process including preparation and training. Lastly, reflexive monitoring relates to participants' ability and intentions to perform formal or informal appraisals of the interventions (May, 2006; May & Finch, 2009).

The study received ethical clearance as stipulated by the Cape Peninsula University of Technology as well as the department of health in the Eastern Cape province of South Africa and the hospital. Informed consent was also sought from all the individual participants of the study.

## 4. RESULTS: INTERVIEW DATA

As mentioned in the previous section, one of the objective of the interviews was to determine participant's awareness and use of eHealth technologies at the hospital. This was used to determine factors that may influence the integration of eHealth technologies with healthcare work activities. To determine the status of use, the researchers during the interview process, searched for indications of awareness and understanding of the purpose of eHealth technologies and its use. The researchers also examined whether the systems were useful in achieving the healthcare practitioner's work objectives, level of use or non-use of the systems and whether the respondents found the system easy to use or not.

In terms of awareness and use of the eHealth technologies at the hospital, participants both in the administrative and clinical process, indicated their awareness and understanding of the purpose of eHealth technologies. There was a consensus among the respondents that eHealth technologies include ICT tools such as hardware, software, networks, and mobile phones that they used to carry out their work activities. Respondents used the terms 'system (s)' and 'technology' interchangeably, therefore the researcher took into consideration such factors while asking the questions. For example, one respondent in the clinical process noted "yes, sometimes we use these systems to order patients' blood tests and receive the results on our mobile phones" (DRG). Respondents within the administrative process gave clear descriptions of their knowledge of existing systems, including Delta 9™, Rx Solution, PAC system and the laboratory systems (PRS-M; PRJ-M; TP.ITS; NM-EC) that were being used at the hospital.

The respondents also noted the purpose of the systems and had positive perceptions of their usefulness and benefits. Findings from the respondents at the hospital revealed that all groups of participants perceived that eHealth technologies, regardless of how they perceived them were or would be useful in their work activities. For example, respondent (RN-N) states that "…I insisted that we type our theatre slates …in order for us to always have our softcopy back up because folders get lost…" In the administrative process, one of the common response in terms of the purpose of eHealth technologies was to improve the management of patient registration data at the hospital (TP.ITS; PRS-M). In another case, respondents in management position indicated that in the laboratory unit at the hospital, the electronic gate keeping system was integrated to ease the burden of cost in that unit that had skyrocketed over the years. Some respondents in the clinical process were unaware of what direct impact the eHealth technologies had on their work activities. As one respondents states "I honestly do not see the benefit of using these technologies in my line of work…" (DCL).

These conflicting views may be attributed to the fact that the hospital lacks adequate support in managing some eHealth technologies. This is coupled with the "lack of interconnectedness" (TP.ITS) of the existing technologies, and some cases "lack of connectivity of some tools" (RN-N) to the hospital network. One respondent mentions that sometimes there was a lack of follow up on the systems, "…there was a software here at the hospital to view x-rays it stopped working at has never been repaired…so we are forced to seek other alternatives" (RN-N).

Table 1 presents the emerging findings on factors that influence eHealth technology use at the hospital.

| **Theme** | **Categories** |
|---|---|
| Technical factors | ✓ Functionality of the system <br> ✓ User skills/competencies <br> ✓ Usability of the system <br> ✓ Constant availability, accessibility and relevance <br> ✓ Embeddedness of work activities into the structure of eHealth technologies |
| Institutional factors | ✓ Advocacy for use and support <br> ✓ Adequate change management <br> ✓ Clarity of system's purpose in the workflow process |
| Behavioral (human) factors | ✓ Reluctance to use <br> ✓ Perceived usefulness of the system <br> ✓ Perceived benefit/value of the system in work activities <br> ✓ Perceived effectiveness of the system in the workflow processes |

**Table 1: Factors that influence the use or non-use of eHealth technologies at the hospital**

Findings reveal that these issues, in most cases, deter participants from making use of the systems, especially those that already see no value in the systems for their work activities. The researchers observed that these issues led to views that the presence of eHealth technologies was disruptive, with one respondent in the clinical process stating "where will you place this technology? It will be in the way" (DCL). The findings show that there is a high level of awareness and relative acceptance of eHealth technologies by administrative staff; however, this is relatively low for clinical staff at the hospital. The high level of awareness and acceptance are attributed to the positive performance expectancy associated with the perceived value of using these technologies in their work activities. Findings also reveal that although there is limited use of the technologies among the healthcare practitioners in the clinical process, there was evidence of a willingness to use. The non-use was attributed to the limited or unreliable ICT infrastructure (such as non-functioning computers, and poor network connectivity at the hospital). Table 2 presents the dominant themes from the finding on the integration of eHealth technologies at the hospital.

| **eHealth technology integration outcomes** | |
|---|---|
| **Theme** | **Description** |
| ✓ Improved turnaround times in the delivery of care services | As a result of integration of eHealth technology such as the laboratory information system, the hospital has seen improved turnaround times in the results of lab tests. This has also seen duplication of data |
| ✓ Improved decision-making and reporting process | The DHIS system has enabled the timely movement of clinical information across the hospital and the public health sector for informed decision making on aspects such as resource allocation or management of disease outbreaks, timely health information needs to reach the relevant authorities. |
| ✓ Lack of systematic integration plan | The hospital has quite a number of eHealth technologies integrated into work activities, however, none of the relevant participants could give a systematic process for integration |
| ✓ Duplication of healthcare data | Despite eradicating duplication of data in some units the eHealth technologies also create duplication of data and process because most systems are not integrated together resulting in duplication of data across the hospital |
| ✓ Poor coordination of existing ehealth technologies (leading to fragmentation) | Most of the systems implemented in a silo, ad hoc manner. The eHealth technologies at hospital are integrated for specific tasks but are not integrated with existing systems |
| ✓ Scalability issues | Some systems like the Delta9 system used at the hospital do not allow flexibility ( although there were new modules of the system available, the system had not be upgraded or scaled to meet the needs in the healthcare service delivery process) |

**Table 2: Themes on the outcomes of eHealth technology integration at the hospital**

## 5. Limitations of the study

A key limitation of this paper is the use of a single-case study. Literature has criticized the limitations of case study strategies, which include lack of ability to generalize the findings, perceived inadequate rigour in case study research, and so forth. The paper acknowledges these critiques, the intentions are not to generalize the findings but to use the outcomes as a starting point in the integration of eHealth in healthcare settings.

## 6. ANALYSIS AND DISCUSSIONS

In this section, the discussions center on elucidation of the findings reported in the previous section. They draw on the key elements of the ActAD model as a theoretical analytical tool. The explanations and causal analysis are based on the findings from the objectives of this study and are done within the critical realist paradigm.

To frame the study's theoretical stance, the researchers were guided by the five basic underlying principles of the ActAD model (Korpela et al., 2004). The first principle is that of an object-oriented activity system, in this study, the eHealth technology integration process is taken as the activity system and the prime unit of analysis. The second principle, describes the activity system as having multiple actors, in this case, the actors include all the relevant stakeholders involved in the eHealth integration process including healthcare practitioners, policy makers, leaders and managers at all levels of the healthcare system. The third principle describes an activity system as emerging as a result of historical activities that are typically formed over a period of time. The use of technologies in the South African healthcare system has seen various historical activities such as the integration of DHIS. The forth principle scrutinizes the fundamental role of contradictions in the activity system as sources of change and development. The fifth principle refer to the possibilities of transformation and the reconceptualisation of the objects and motives in the activity system (Engeström, 2001). The study interprets this as the transformation brought about by inhibiting and enabling factors in the activity system. These factors have the ability to transform the manner in which activities are carried out in the activity system. The outcomes of the transformation process may be desired or undesired. A closer investigation of these principles reveals that they possess similar traits of the critical realism paradigm. The principles are relatively broad with regard to the methods of application. They provide an overarching frame and conceptual tools of enquiry; in essence the principles provide exploratory guidance rather than rigid rules.

The discussions in this section are based on the outcomes (Table 2) of the eHealth technologies integration efforts at the empirical case. As defined in the introductory section of the paper, integration is defined as the process of putting to meaningful use technology interventions within a healthcare setting to achieve the goals improved healthcare service. The study argues that context-based factors presented in Table 1 determine use or non-use of eHealth technologies at the facility. Drawing on the ActAD model's fifth principle, the context-based factors informs and transforms the work activities in the integration of eHealth technology work activity system to produce the outcomes in Table 2. Addressing such factors may lead to enhanced integration of eHealth technologies with healthcare work activities for an improved healthcare information collection, processing and reporting. The study argues that the outcomes of the current eHealth integration practices, are dependent variables that are mediated by context-based factors, which in turn transform eHealth integration activities. The outcomes of this transformation process generate either observable or unobservable events in the delivery of healthcare services. Guided by the fourth principle of scrutinizing the presence of contradictions, the study deduces that the outcomes allude to the presence of inadequacies in the current integration practices despite success in other areas. This contradictions, drive the changes in the healthcare service delivery process. Figure 1 depicts the interrelationship between the independent and dependent variables in the eHealth technology integration activity system.

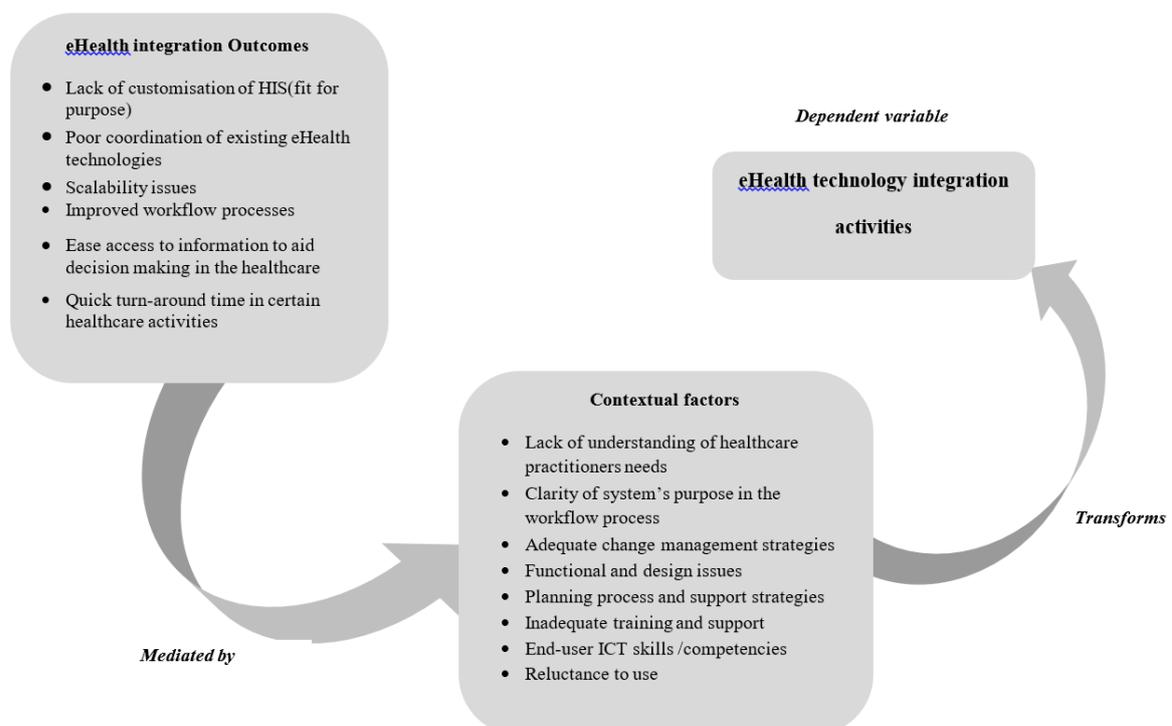

**Figure 1: The interrelationship between context-based factor and eHealth integration outcomes**

## 6.1. Discussions on the context-based factors that influence ehealth technology integration

In the following sub-section, the context-based factors that transform the eHealth technology integration activities.

i. Clarity of system's (eHealth technologies) purpose in the workflow process

A crucial factor that transforms activities in the integration process of eHealth technologies in healthcare settings is the need for clarifying eHealth purpose in the healthcare workflow process. Integration of eHealth in healthcare facilities requires a restructure of workflow processes to accommodate the technologies. It is therefore vital that to understand how eHealth technologies not only align with existing work activities, but also the value they bring to the workflow. Without clear articulation of the need for and purpose of eHealth technologies, in most cases result in sub-optimal use and fragmentation of healthcare services as suggested by some participants in the empirical case (TP.ITS; NM-EC). For a successful integration of eHealth technologies to be realized, the functions of the technologies must align with and form an integral part of the healthcare workflow processes (Eder & Igbaria, 2001).

Another implication of this factor is that without clear articulation of system purpose, is likely to lead to healthcare practitioners' resistance to use of the systems, as healthcare practitioners would perceive this as extra work. eHealth integration success is dependent on the user's commitment to use the technology (Holahan et al., 2004). As postulated by Ynalvez and Shrum (2011), the work environment may improve; however, the effects on work activities depend on how individuals make use of the systems.

ii. Change management strategy
Another factor is the need for change management strategy (s), this involves the management and organizing the activities of eHealth integration process, managing expectations of users, training healthcare practitioners and outlining the roles and responsibilities in the integration process. The absence of this elements during the integration process can influence and transform the activities into the observed outcomes. The findings reveal that part of change management strategy is ensuring appropriate resources are allocated and made available in a coordinated manner
This implies that the environment has to be conducive to enable optimal integration of eHealth technologies with healthcare work activities. Another aspect of the change management factor is inform all stakeholders of the values, benefits and purpose of ehealth technologies in the healthcare workflow. Adequate change management strategies all for planning for activities such as planning for assessment of the healthcare setting for relevant adequate resources such as ICT and networking infrastructure, equipping healthcare practitioners with the right ICT skills. This, coupled with factors such as political will from regional managers (DMA) to support the process long term would create an initial facilitating environment for eHealth integration.

iii. Technical and Organizational factors

Technical factors such as functionality of the system influence the use of ehealth technologies. A reliable system improves the confidence of end-users in pursuit of operational objectives. As one participant from the hospital noted, "The hospital experiences a lot of network problems … [as a result] we do get a lot of frustrated users" (TP.ITS). User skills/competencies also influence use or non-use of eHealth technologies at the hospital. User competencies here imply the understanding, literacy and ability to put technologies into effective use. In addition, technical factors include uncoordinated systems at the hospital that force participants to duplicate processes. As mentioned by one participant: "If we could have systems that are connected to the network so that you just fill in patient information and you store it so that even for your referrals your colleague from another hospital just needs to punch in the folder number and see what was done" (RN-N). Poor technical support of networked systems is also an inhibitor of HIS use: when the support takes long to reach the end user, frustration leads to non-use.

These context-based factors are triggered by underlying generative mechanisms discussed in the next sub-section.

## 6.2. A discussion of the generative mechanisms in the eHealth technologies integration with healthcare work activities

The discussions in this sub-section centers on the identification, characterization and explanations of generative mechanisms that shape the outcomes of eHealth technologies integration activities. For that reason, the approach in the discussions shifts from ontological to epistemological, where the relationship between mechanisms, events and empirical experiences is stressed (Bhaskar, 1979). NPT categorizes all the work activities in the implementation process according to four main interactive constructs (May & Finch, 2009). According to the authors, although the constructs and their components describe different types of 'work' activities, they are correlated. What this means is that the constructs and the components constantly interact, with the potential to influence and change one another.

The first generative mechanisms that is identified is ***lack of analysis of eHealth fit-for-purpose***. This mechanism is considered as a major deterrent to the integration of eHealth technologies with healthcare work activities. This mechanism, can be categorized under the construct coherence as it relates to how work activities that define and organize the eHealth technology integration with healthcare work activities are understood and perceived meaningful. Findings reveal that there are cases at the hospital where healthcare practitioners such as doctors do not perceive the technologies to have any value to their work activities and as such, see no point in using the technologies. In some instances the practitioners view these technologies as disruptive. Based on such views, the researchers deduce that the underlying mechanism that could trigger such views is the lack of analysis of how the eHealth technologies would fit with the work activities of the healthcare practitioners. The result of this is having the paper-based system still being used even for activities that have supposedly been automated. This defeats the purpose of improving healthcare activities and enabling ease of access to updated information at points-of-care by healthcare practitioners. This mechanism has causal powers that produce outcomes such as lack of system customization to fit healthcare practitioners' work activities, and poorly integrated information systems into the work practices of healthcare practitioners. All these outcomes inhibit HIS implementation for public service delivery.

The second generative mechanism identified is the ***absence of considerable and coordinated infrastructure resources.*** The integration of eHealth technologies with healthcare work activities require major structural reforms in the healthcare system that include provision of adequate distribution of resources, increasing capacity of skilled human resource, provision of funding and so forth. A resource-constrained environment influences how actors in the work activity system carry out their activities. As the findings portray, poor network infrastructure at the hospital influence successful integration of eHealth technologies with healthcare work activities. Unequal funding mechanisms between the public and the private healthcare sector is also an underlying mechanism in the integration of eHealth technologies. This is in terms of how the resources are acquired and maintained sustainably.

In terms of skilled human resource, ***a lack of continuous re-skilling*** of healthcare professionals with the necessary ICT skills, also deters the integration process. The implications of this mechanism manifests in inaccurate data capturing that leads to repetitive processes and therefore long turnaround times. Findings also revealed that professional skills and competency affect work attitudes, including commitment from top management.

The outcomes of these mechanisms are contingent on other context-based mediators such as the widespread socio-economic challenges the country faces, discrepancies in translating health policies/strategies into practice and so forth. From the analysis, the study observed that the eHealth integration outcomes directly or indirectly experienced may be dependent on the complexity of the multiple generative mechanisms. An understanding of the interplay between observed events, structures, conditions and generative mechanisms may explain why and how eHealth technologies integration does not adequately facilitate healthcare service delivery in public healthcare facilities in resource-constrained environments.

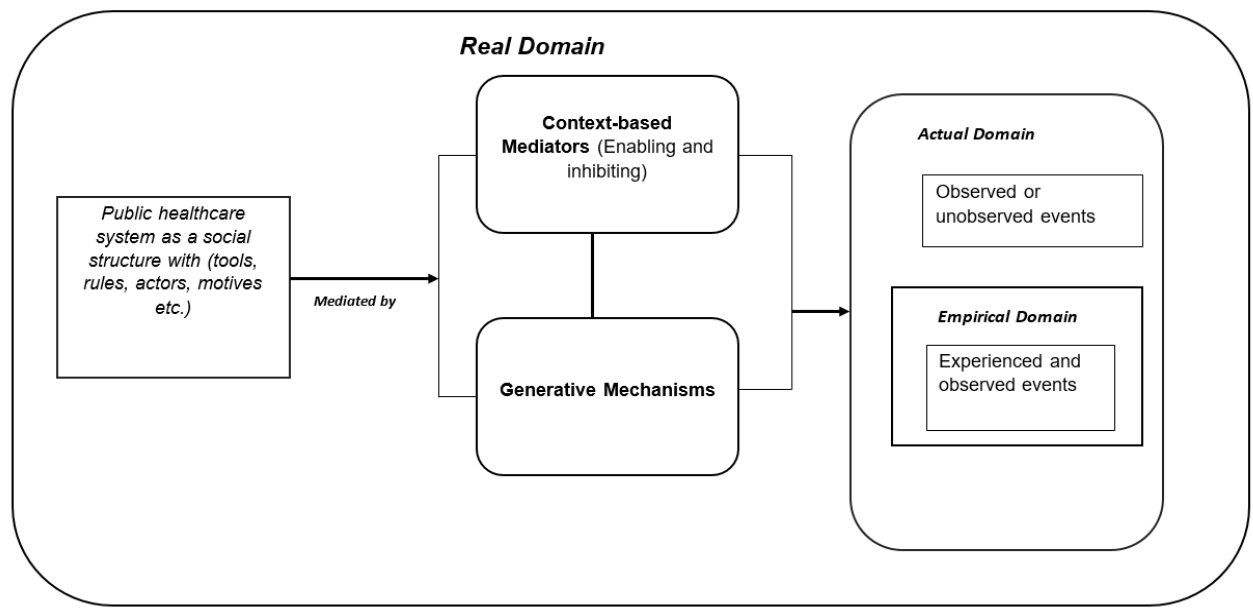

**Figure 2: A stratified representation of how events occur in the HIS integration within a healthcare service delivery context**

**Figure 2** depicts a stratified representation of how events occur in the integration of eHealth technologies with healthcare work activities. The social structures are mediated by context-based factors and generative mechanisms. The generative mechanisms then evoke the events in the actual domain and experiences in the empirical domain. In the actual domain, events experienced include poorly integrated system that hinder work flow in the healthcare setting. In the empirical domain, healthcare practitioners either perceive eHealth technologies as useful or not depending on their experiences with using the technologies. Ultimately, this results into how tasks are carried and how healthcare practitioners perceive their performance in the healthcare service delivery system.

## 7. CONCLUSION

Integration of eHealth technologies with healthcare work activities requires more than understanding and addressing context-based factors that may influence the integration activities. In addition to this, it is crucial to also identify, characterize and explain the underlying generative mechanisms that have causal effects that produce the outcomes in the current integration efforts of eHealth technologies with healthcare work activities. It is based on this argument and the objectives of this paper that the analysis of the empirical data was done. The analysis in the end reveals the factors such as a need for clarity of ehealth technology's purpose in the healthcare workflow mediate integration outcomes such as the lack of customised technologies to fit healthcare work activities. The underlying generative mechanism that may trigger such events include the absence of analysing systems for fit for purpose.

The paper also reveals that generative mechanisms such as the absence of continuous re-tooling of healthcare practitioners with ICT skills can help explain the some of the outcomes of the current integration efforts of eHealth technologies with healthcare work activities in public healthcare facilities. The findings in this paper presents a critical analysis can be of benefit for the implementer and relevant government authorities in the healthcare sector. Another contribution the paper makes can be used to better guide future eHealth technology integration with healthcare work activities in different healthcare settings. The paper has provided reasons why integration of eHealth technologies do not adequately facilitate healthcare practitioners work activities. These reasons are due to the contextual and due to the contradictory interplay between social and technical generative mechanisms.

The study proposes that future work on eHealth integration in healthcare contexts should focus on developing frameworks for assessing the sustainability of eHealth in the public healthcare space and also on assessing the process of operationalizing health policy/strategy at facility level to evaluate the impact, benefits and value of these policies and strategies on the healthcare system.